\documentclass[10pt,conference]{IEEEtran}
\IEEEoverridecommandlockouts
\usepackage{cite}

\usepackage{hyperref}
\usepackage{amsmath,amssymb,amsfonts}
\usepackage{algorithmic}
\usepackage{graphicx}
\usepackage{subfig}
\usepackage{textcomp}
\usepackage{xcolor}
\usepackage{fancyhdr}
\def\BibTeX{{\rm B\kern-.05em{\sc i\kern-.025em b}\kern-.08em
    T\kern-.1667em\lower.7ex\hbox{E}\kern-.125emX}}

\fancypagestyle{FirstPage}{
\lfoot{979-8-3503-4323-6/23/\$31.00 \copyright 2023 IEEE \\
DOI 10.1109/QCE57702.2023.10211}}

\begin{document}

\title{Resource Estimation of Quantum Multiplication Algorithms\\
}

\author{\IEEEauthorblockN{Ethan R. Hansen}
\IEEEauthorblockA{\textit{Department of Physics} \\
\textit{University of Washington}\\
Seattle, USA \\
ethanrh@uw.edu}
\and
\IEEEauthorblockN{Sanskriti Joshi}
\IEEEauthorblockA{\textit{Department of Electrical and Computer Engineering} \\
\textit{University of Washington}\\
Seattle, USA \\
sjoshi3@uw.edu}
\and
\IEEEauthorblockN{Hannah Rarick}
\IEEEauthorblockA{\textit{Department of Physics} \\
\textit{University of Washington}\\
Seattle, USA \\
rarichan@uw.edu}
}

\maketitle

\begin{abstract}
As quantum computers progress towards a larger scale, it is imperative that the “top” of the computing-technology stack is improved. This project investigates the quantum resources required to compute primitive arithmetic algorithms, particularly multiplication. By using various quantum resource estimators, like Microsoft’s Azure Quantum Resource Estimator, one can determine the resources required for numerous quantum algorithms \cite{b5}. In this paper, we will provide a comprehensive resource analysis of numerous quantum multiplication algorithms such as Karatsuba, schoolbook, and windowed arithmetic for different qubit platforms (trapped ion, superconducting, and Majorana) using the new Azure Quantum Resource Estimator.
\end{abstract}


\section{Introduction}
\thispagestyle{FirstPage}

As the hardware and software for quantum computers are designed, focus should be placed on optimizing all levels of the computing technology stack to ensure a maximally efficient quantum computing system. The promise of quantum computers is its ability to solve large P and find approximate solutions for NP problems in a much more time-efficient manner than its classical counterpart \cite{b1a}. Integer factorization is an example of this. A major component of RSA decryption is factoring a number that is a product of two large primes. This is computationally intensive, but, using Shor's algorithm, there is a possibility of the factorization taking a fraction of the time on a quantum computer \cite{b7}. However, solving a large problem would be resource intensive, requiring many orders of magnitude of the number of qubits than what currently exists in a system. Additionally, qubits tend to be noisy, and at that scale would require a form of error correction \cite{b1a}. To implement these different levels of algorithms, an analysis of the resources required is crucial for understanding the requirements for hardware based on software implementation \cite{b1}.

To date, researchers are investigating transforming classical algorithms for arithmetic, such as multiplication, into quantum algorithms to be used on quantum processors\cite{b2, b22}. In particular, quantum multiplication algorithms will be needed for RSA applications and Shor's algorithm. Although there are publications related to the various quantum algorithms for multiplication, none of them provide an analysis of the resource estimation using Microsoft Azure Quantum Resource Estimator which accounts for quantum error correction and provides insights into algorithm runtime and the physical qubits required \cite{b3}\cite{b4}. Resource estimation provides different metrics to understand the approximate resources required to run a quantum algorithm on a specific hardware setup, including the number of qubits, quantum gates, time, etc.\cite{b5}. Understanding how many resources would be required to run an algorithm helps define a scale for how large and resilient to noise a quantum computer has to be to effectively run the algorithm \cite{b55}.

In the subsequent sections, we will introduce the multiplication algorithms used in this project (Sec.\ref{Multiplication Algorithms}), discuss our approach to resource estimation of these algorithms using Azure Quantum Resource Estimator (Sec.\ref{methods}), and present an analysis and conclusion of our findings (Sec.\ref{results}-\ref{conclusion}). 
\section{Multiplication Algorithms}\label{Multiplication Algorithms}
This project will focus on determining the resource estimation for multiplication primitives. Two different algorithms were used in addition to schoolbook multiplication: Karatsuba multiplication and windowed multiplication \cite{b3, b4}. While these algorithms are classical, it is non-trivial to apply them on a quantum computer. To convert such an algorithm, the operations cannot be irreversible as it can cause decoherence, whereas large quantum algorithms require coherence \cite{b3}.

Schoolbook multiplication is the method taught in grade school for long multiplication. This multiplication requires $\mathcal{O}(n^{2})$ operations.  
\subsection{Karatsuba multiplication}
Karatsuba multiplication requires a sub-quadratic number of operations by recursively multiplying, while using those results to calculate the final answer. For example, to perform the multiplication of integers $a$ and $b$, each integer gets broken up such that $a=c+2^{h}d$ and $b=e+2^{h}f$. Using the results of recursively multiplying $ce, df, (c+d)(e+f)$ yields the desired result \cite{b3}. In addition, for the quantum case, the coherence is gained by having the intermediate values added directly to sections of the output register \cite{b3}. Notably, this method of removing decoherence requires the same $\mathcal{O}(n)$ space usage and $\mathcal{O}(n^{lg3})$ operation count as the original classical algorithm.

\subsection{Windowed multiplication}
Windowed arithmetic is implemented to reduce the number of operations counts using look-up tables to merge together operations. These look-up tables are classically precomputed. Additionally, measurement based uncomputations of the look-up tables can be done to gain coherence \cite{b4}. This method has asymptotic Toffoli gate count of $\mathcal{O}(\frac{n^{2}}{lgn})$. 

\section{Methods}\label{methods}
We will be working with Microsoft’s quantum resource estimation tool to investigate and compare the resources required for different integer multiplication algorithms. The main estimation tool that we will utilize is the Azure Quantum Resource Estimator. 

The Azure Quantum Resource Estimator (cloud resource estimator) is run on Microsoft's machines and is accessed on the cloud via the Azure Portal. The cloud resource estimator takes in three inputs: the quantum circuit, the desired physical qubit parameters (error rates, gate times, and measurement times), and the quantum error correction scheme to utilize \cite{b1}. At the logical level, the estimator compiles the provided circuit down to a universal, fault-tolerant gate set required for implementation on a quantum computer \cite{b1}. Given the provided qubit error rates and the quantum error correction scheme, the estimator calculates how many physical qubits it will take to form a single logical qubit \cite{b1}. 
The estimator formats the circuit to account for the 2D connectivity requirements of current quantum processor technologies. Given that two-qubit gates and qubit entanglement are achieved through local qubit interactions, a 2D array of qubits makes it difficult to perform operations on one side of the processor with a qubit on the other side of the processor without any intermediate operations. To account for this 2D constraint, the cloud resource estimator assumes a fast block layout scheme where clusters of two algorithm qubits are surrounded by a ring of ancilla qubits such that a path from any two qubits on the quantum processor can be formed via these ancilla qubits, thus allowing for any arbitrary Pauli operation to be performed \cite{b55}. This process provides a more realistic estimate for the number of logical qubits, both algorithm and ancilla qubits, required to physically implement any algorithm on a 2D quantum processor. 
The fast block layout maps a logical circuit with $Q_{alg}$ logical qubits to Q total logical qubits according to the Eq. \ref{Qalg to Qtotal} shown below \cite{b1}:
\begin{equation}
\label{Qalg to Qtotal}
    Q =2 Q_{alg}+\lceil \sqrt{8Q_{alg}} \rceil +1 
\end{equation}
Estimates for the required logical and physical resources (qubits, T-states, runtime) are then calculated based upon the formatted circuit, the number physical qubits to form a logical qubit, and the gate and measurement times \cite{b1}. 

Six presets for qubit paramters are provided based on three different qubit platforms, gate-based superconducting and trapped-ion systems and Majorana systems which have yet to be implemented. There are two presets for each platform, one realistic estimate and one optimistic estimate for the future gate times and error rates of a fault-tolerant quantum processor based upon that qubit platform \cite{b1}. Currently, two quantum error correction models are supported, the surface code model, which is used by default and works with all qubit platforms, and the floquet code model, which is designed for and only works with Majorana systems.

In Sec. \ref{subsec:complexity}, we will use the default qubit platform (gate-based quantum processor with 50 ns gate times, 100 ns measurement times, and $10^{-3}$ single- and two-qubit gate errors) and the default surface code error correction scheme to investigate the physical-space complexity and the time complexity \cite{b1}. 

In Sec. \ref{subsec:platform}, we will investigate estimated physical qubits and algorithm runtime for performing plus-equal multiplication of random 2048-bit integers across the platform presets provided by the resource estimator. For error correction, we will use the floquet code for the Majorana platform as it is more efficient than the surface code \cite{b1}. We will use the surface code for all other platform presets on the cloud resource estimator. 

To date, literature on resource estimation for multiplication algorithms uses the QCTraceSimulator implemented in Microsoft's Quantum Development Kit (QDK) for Q\# \cite{b3, b4}. In this project, we will refactor these algorithms utilizing the new Microsoft resource estimation tool (Azure Quantum Resource Estimator). Then, taking advantage of metrics allotted by the cloud resource estimator (algorithm runtime and required physical qubits), we will investigate whether any of the multiplication algorithms can provide a significant advantage over the schoolbook algorithm.

\section{Results}\label{results}
\label{sec:Res}

All results in this paper will be from cloud estimator as the local estimator results are already given in the literature \cite{b3}. We have verified that the logical results from the cloud resource estimator match up with the results from the local resource estimator after applying Eqn. \ref{Qalg to Qtotal} to the local estimators data to account for the 2D layout processing of the cloud estimator. 

\subsection{Physical-Space, Time, and T-State Complexity}
\label{subsec:complexity}
\begin{figure}[h]
    \centering
        \subfloat{\includegraphics[scale=0.42]{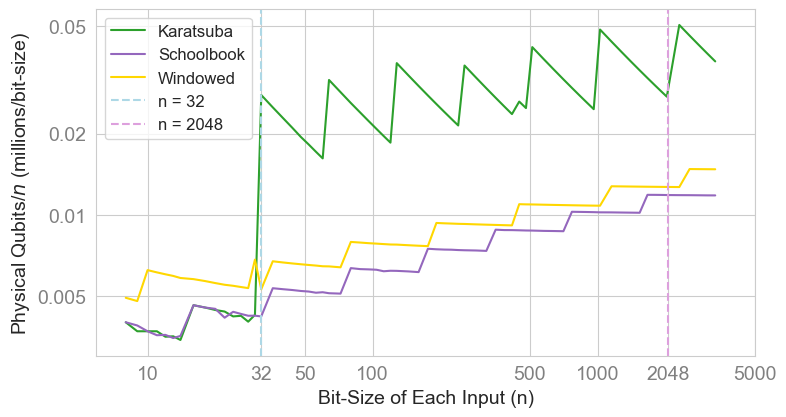}}
        \qquad
        \subfloat{\includegraphics[scale=0.42]{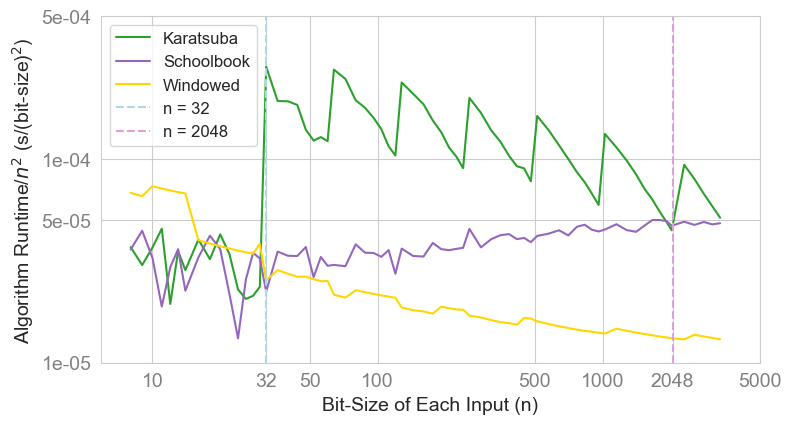}}
        \qquad
        \subfloat{\includegraphics[scale=0.42]{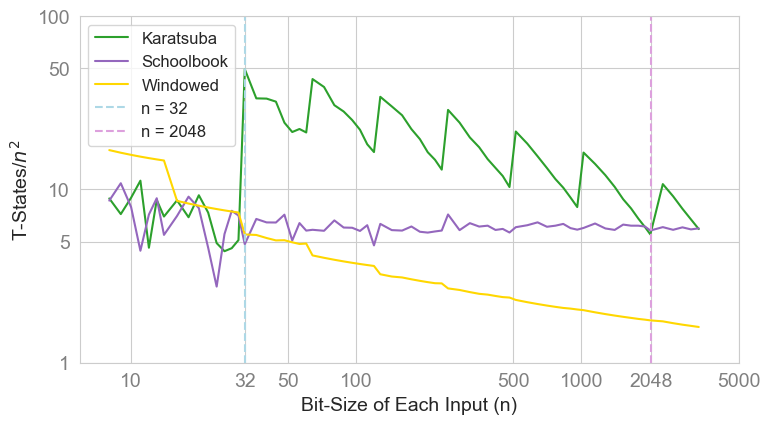}}
        \caption{Estimated resources required for performing plus-equal multiplication ($a+=b*c$) using the cloud resources estimator (default gate\_ns\_e3 platform using the surface code quantum error correction). (Top) Estimated physical qubits (millions) divided by the bit-size of each input (a, b, and c are random n-bit numbers). (Middle) Estimated algorithm runtime (seconds) divided by $n^{2}$. (Bottom) Estimated T-states (millions) divided by $n^{2}$.} 
    \label{fig:Time_Complex}
\end{figure}

 We analyze the physical-space complexity of the Karatsuba, schoolbook, and windowed algorithms in the top plot of Fig. \ref{fig:Time_Complex}, where the vertical axis is the physical qubits (in millions) divided by the bit-size of the inputs, n. We observe that schoolbook multiplication utilizes the least amount of physical qubits for all factor sizes plotted. The windowed and schoolbook algorithms exhibit a stair-step behavior in physical qubits with regions of constant qubits/bit-size separated by discrete jumps at select bit-sizes. While the discrete jumps are correlated between the schoolbook and windowed algorithms at lower bit-sizes, the jumps for the schoolbook algorithm begin to increasingly occur at lower bit-sizes than windowed as bit-size increases, suggesting that schoolbook could eventually surpass windowed in required physical qubits.

We analyze the time complexity of the three algorithms in the middle plot of Fig. \ref{fig:Time_Complex}, where the vertical axis is the algorithm runtime (in seconds) divided by the square of n. We note that, for the case of the windowed algorithm, the algorithm runtime does not include the time required to classically compute the look-up table, however this process can be done prior to running the algorithm and saved to be used for all future multiplications. For bit-sizes greater than 32, the windowed algorithm performs faster than the other two algorithms. The Karatsuba algorithm begins to gain an advantage over schoolbook at 2047-bit inputs, and will eventually surpass windowed for large inputs, much larger than the current sizes utilized in RSA procedures.

The T-state complexity, shown in the middle plot of Fig \ref{fig:Time_Complex}, follows a similar trend as the time complexity. This correlated behaviour is to be expected given the much larger time-cost of producing high fidelity T-gates/T-states compared to Clifford gates. The T-state complexity scales as $\mathcal{O}\left(n^2\right)$, while the time complexity has some non-trivial scaling for lower bit-sizes, but appears to converge to $\mathcal{O}\left(n^2\right)$ scaling for the larger bit-sizes shown. This is consistent with the logical scaling shown in the literature \cite{b4}.

\begin{figure*}
    \centering
    \includegraphics[width=\textwidth]{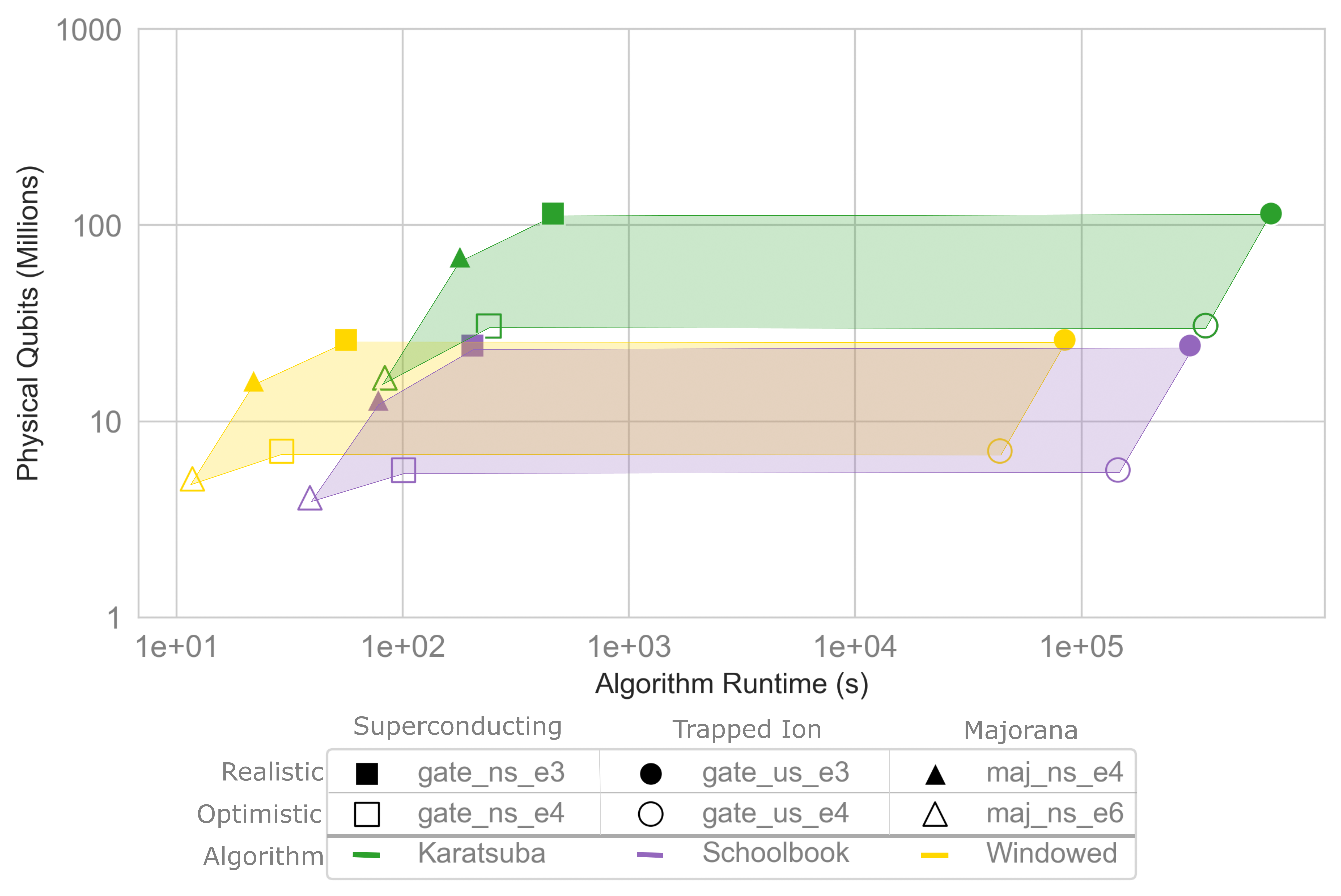}
    \caption{Estimated physical qubits and algorithm runtime across different platforms for plus-equal multiplication with 2048-bit inputs. For the Majorana system, the floquet code is utilized for error correction, while the other platforms use the standard surface code.}
    \label{fig:Platform}
\end{figure*}

\subsection{Estimated Resources Across Qubit Platforms}
\label{subsec:platform}
We now analyze how the resources for each algorithm compared across different platforms. We investigate the three qubit platforms provided by the Azure Quantum Resource Estimator: superconducting, trapped ion, and Majorana systems. For each system, we consider a realistic platform with realistic gate/measurement times and errors and a hopeful platform with optimistic realistic gate/measurement times and errors (see \cite{b5} for more on custom resource estimation parameters). We show the algorithm runtime and required physical qubits for each algorithm, using a factor size of 2048 across the six different platforms, in Fig. \ref{fig:Platform}.

As expected from Fig. \ref{fig:Time_Complex}, we see that the windowed algorithm utilizes slightly more physical qubits than schoolbook; however, the algorithm runtime is shorter for windowed for this input size. The difference between windowed and schoolbook is minimal on the optimistic Majorana system, 5 million qubits for windowed compared to 4 million qubits for schoolbook on the optimistic Majorana system. This difference is even more negligible for the realistic superconducting platform, 26 million qubits for windowed compared to 24.2 million qubits for schoolbook. The Karatsuba algorithm doesn’t stack up well against the other algorithms for 2048-bit numbers, but, as discussed in Sec. \ref{subsec:complexity}, Karatsuba has the potential to utilize less T-states overall for larger bit-sizes. 

We  observe that all algorithms perform more efficiently on the Majorana system, supporting the results from \cite{b1}. 

\section{Conclusion and Outlook}\label{conclusion}


Resource estimation is crucial for understanding the resources required for various quantum algorithms. Although quantum algorithms discussed in this paper required far more resources than currently possible, it is still imperative to understand how the resource requirements will scale for algorithms as quantum computers grow to larger scales. Azure Quantum Resource Estimator provides an alternative route to resource estimation of arithmetic primitives. Using the Azure Quantum Resource Estimator, we were able to show the physical-space, time, and T-state complexity for the Karatsuba, windowed, and schoolbook multiplication algorithms. We find that the time complexity and the T-state complexity has the same scaling, suggesting that the algorithm run time is dominated by performing these T-gates. Furthermore, when all three algorithms are compared on the different platforms (trapped ions, superconducting, and Majorana), the windowed algorithm performs the best for all three platforms with Majorana being the most efficient. 

These algorithms were initially designed to make classical multiplication more efficient. Therefore, they have yet to take in account how it would be physically implemented on qubits. The algorithms have the potential for further improvement for different qubit platforms by considering physical implementation, which would be the next step in comparing the arithmetic primitive algorithms.

Further information regarding the procedures required to reproduce the results for various quantum algorithms is located via a public Github repository (\url{https://github.com/hdrarichan/UW\_EE522\_SP2023}).

\section*{Acknowledgements}
We would like to acknowledge Dr. Wim van Dam and Mariia Mykhailova for guidance on this project. Additionally, we would like to acknowledge Prof. Sara Mouradian, Prof. Boris Blinov, and the Accelerating Quantum Enabled Technologies program at the University of Washington for coordinating this research. 


\end{document}